\documentclass{sf2a-conf2013}
\usepackage{graphicx}
\usepackage{hyperref}
\usepackage[]{natbib}  
\usepackage{epstopdf}

\def\BibTeX{{\rm B\kern-.05em{\sc i\kern-.025em b}\kern-.08em
    T\kern-.1667em\lower.7ex\hbox{E}\kern-.125emX}}
\bibpunct{(}{)}{;}{a}{}{,}  

\begin{document}

\TitreGlobal{SF2A 2013}


\title{World-leading science with SPIRou - \\
The nIR spectropolarimeter / high-precision velocimeter for CFHT 
}

\runningtitle{Science with SPIRou}

\author{X. Delfosse}\address{UJF-Grenoble 1/CNRS-INSU, Institut de Plan\'etologie et d'Astrophysique de Grenoble (IPAG) UMR 5274, 38041 Grenoble, France}

\author{J.-F. Donati}\address{UPS-Toulouse / CNRS-INSU, Institut de Recherche en Astrophysique et Plan\'etologie (IRAP) UMR 5277, Toulouse, F31400 France}

\author{D. Kouach$^2$}

\author{G. H\'ebrard}\address{Institut dAstrophysique de Paris, CNRS, Universit\'e Pierre et Marie Curie, 98bis Bd. Arago, 75014 Paris, France}

\author{R. Doyon}\address{D\'epartement de physique and Observatoire du Mont-M\'egantic, Universit\'e de Montr\'eal, C.P. 6128, Succursale Centre-Ville, Montr\'eal, QC H3C 3J7, Canada}

\author{E. Artigau$^4$}

\author{F. Bouchy}\address{Aix Marseille Universit\'e, CNRS, LAM
  (Laboratoire d’Astrophysique de Marseille) UMR 7326, 13388,
  Marseille, France}

\author{I. Boisse$^5$}

\author{A.S. Brun}\address{Laboratoire AIM Paris-Saclay, CEA/Irfu
  Universit\'e Paris-Diderot CNRS/INSU, 91191 Gif-sur-Yvette, France}

\author{P. Hennebelle$^{6,}$}\address{LERMA (UMR CNRS 8112), Ecole
  Normale Sup\'erieure, 75231, Paris Cedex, France} 

\author{T. Widemann}\address{Paris Observatory, LESIA UMR 8109,
  Meudon, France}

\author{J. Bouvier$^1$}

\author{X. Bonfils$^1$}

\author{J. Morin}\address{LUPM, Universit\'e Montpellier II, CNRS, UMR
  5299, Place E. Bataillon, 34095, Montpellier, France}

\author{C. Moutou}\address{Canada-France-Hawaii Telescope Corporation,
  65-1238 Mamalahoa Hwy., Kamuela, HI 96743, USA}

\author{F. Pepe}\address{Observatoire Astronomique de l'Universit\'e de Gen\`eve, 51 Ch. des Maillettes, 1290 Sauverny, Versoix, Suisse}

\author{S. Udry$^{11}$}

\author{J.-D. do Nascimento}\address{Departamento de F\'{\i}sica
  Te\'orica e Experimental (DFTE), Universidade Federal do Rio Grande
  do Norte (UFRN), CP 1641, 59072-970 Natal, RN, Brazil}

\author{S.H.P. Alencar}\address{Departamento de F\'{\i}sica - ICEx - UFMG,
  Av. Antônio Carlos, 6627, 30270-901 Belo Horizonte, MG, Brazil}

\author{B.V. Castilho}\address{Laborat\'orio Nacional de
  Astrof\'{\i}sica/MCT, Rua Estados Unidos 154, 37504-364 Itajub\'a, MG,
  Brazil}

\author{E. Martioli$^{14}$}

\author{S.Y. Wang}\address{Institute of Astronomy and Astrophysics,
  National Taiwan Univ., Taiwan}

\author{P. Figueira}\address{Centro de Astrof\'{\i}sica, Universidade
  do Porto, Rua das Estrelas, 4150-762, Porto, Portugal}

\author{N.C. Santos$^{16}$}

\author{the SPIRou Science Team}

\setcounter{page}{1}


\maketitle


\begin{abstract}

SPIRou is a near-infrared (nIR) spectropolarimeter / velocimeter
proposed as a new-generation instrument for CFHT. SPIRou aims in
particular at becoming world-leader on two forefront science topics,
(i) the quest for habitable Earth-like planets around very- low-mass
stars, and (ii) the study of low-mass star and planet formation in the
presence of magnetic fields. In addition to these two main goals,
SPIRou will be able to tackle many key programs, from weather patterns
on brown dwarf to solar-system planet atmospheres, to dynamo processes
in fully-convective bodies and planet habitability. The science
programs that SPIRou proposes to tackle are forefront (identified as
first priorities by most research agencies worldwide), ambitious
(competitive and complementary with science programs carried out on
much larger facilities, such as ALMA and JWST) and timely (ideally
phased with complementary space missions like TESS and CHEOPS).

SPIRou is designed to carry out its science mission with maximum
efficiency and optimum precision. More specifically, SPIRou will be
able to cover a very wide single-shot nIR spectral domain (0.98-2.35~$\mu$m) 
at a resolving power of 73.5K, providing unpolarized and polarized spectra 
of low-mass stars with a $\sim$15\% average throughput and a radial velocity (RV) precision of 1~m/s.

\end{abstract}

\begin{keywords}
Extrasolar planets, Super-Earths in the habitable zone, 
Star / planet formation, Stellar magnetic fields, 
Velocimetry, Spectropolarimetry.
\end{keywords}


\section{Introduction}
 
The science programs SPIRou proposes to tackle are forefront (first priorities for most research agencies worldwide), 
ambitious (competitive and complementary with science programs carried out on much larger facilities, e.g., ALMA/ESO and JWST/NASA) 
and timely (ideally phased with complementary instruments, e.g., TESS/NASA, CHEOPS/ESA and JWST/NASA).
SPIRou plans to concentrate on two main scientific goals. The first one is to
search for and characterize habitable exo-Earths orbiting very-low
mass stars (vLMSs) using high-precision radial velocity (RV)
measurements. This search will expand the initial, exploratory studies
carried out with visible instruments (e.g., HARPS/ESO) and will survey
in particular large samples of stars mostly out of reach of existing
instruments. In particular, carrying out a new large- scale survey at
nIR wavelengths will boost the sensitivity to habitable exo-Earths by
typically an order of magnitude on planetary mass (with respect to
existing instruments). SPIRou will also work in close collaboration
with space- and ground- based photometric transit surveys like
TESS/NASA, CHEOPS/ESA and ExTrA\footnote{ExTrA is a recently funded ERC project whose aim is to detect transiting Earth-like planets around M-dwarfs 
from the ground. As a low resolution multiple-object spectrograph, ExTrA will allow extremely accurate photometry in narrow wavelength bands} 
to identify the true planets among the candidates they will discover.

The second main goal is to explore the impact of magnetic fields on star and planet formation, by detecting fields of various types of young 
stellar objects (e.g. class-I, -II and -III protostars, young FUor-like protostellar accretion discs) and by characterizing their large-scale 
topologies. SPIRou will also investigate the potential presence of giant planets around protostars and in the inner regions of accretion discs. 
In particular, this study will vastly amplify the initial exploration surveys carried out at optical wavelengths within the MaPP (Magnetic 
Protostars and Planets) and MaTYSSE (Magnetic Topologies of Young Stars and the Survival of close-in massive Exoplanets) CFHT Large Programs (LPs). 
It will also ideally complement the data that ALMA/ESO has just started collecting on outer accretion discs and dense prestellar cores.
SPIRou will also be able to tackle many additional exciting research topics in stellar physics (e.g., dynamos of fully convective stars, 
weather patterns of brown dwarfs), in planetary physics (e.g., winds and chemistry of solar-system planets), galactic physics (e.g. stellar archeology) 
as well as in extragalactic astronomy.
We detail these goals below, giving in the main cases the typical samples that need to be explored and their observational properties.

\section{SPIRou}

\subsection{Science requirements and instrument concept}

SPIRou is designed to carry out its science mission with maximum
efficiency and optimum precision. More specifically, SPIRou will be
able to cover a very wide single-shot nIR spectral domain (0.98-2.35$\mu$m) 
at a resolving power of 73.5K, providing unpolarized and polarized
spectra of low-mass stars with a $\sim$15\% average throughput and a 
RV precision of 1~m/s.  Table~\ref{tab_requis} list the
main scientific requirements of SPIRou, needed to carry out most
science goals detailled in Sec~3 and 4.

\begin{table}
\begin{tabular}{|l|l|} \hline
Requirement & Value \\ \hline \hline
Simultaneous Spectral Range & full coverage from 0.98-2.35 μm (YJHK bands)  \\
Resolving Power                     & $>$70K (goal 75K)  \\
RV Precision                           & $<$1 m/s (rms)  \\
Polarimetric Performance       & relative precision: better than 2\% (goal 1\%); sensitivity: 10 ppm  \\
Instrument Sensitivity            & S/N$>$100 per 2 km/s pixel in 1 hr at J=12 \& K=11; \\
                                              & bright limit: H$<$3.5 (goal H$<$1) - faint limit: H$\sim$14  \\
Observational Efficiency        & $>$70\% \& $>$90\% for 15 min \& 1 hr visit respectively  \\
Sky Coverage                         & up to airmass 2.5 (zenithal distance 70$^{\circ}$) \\ \hline
\end{tabular}
\caption{Summary of SPIRou scientific requirements}
\label{tab_requis}
\end{table}

\begin{figure}[ht!]
 \centering
 \includegraphics[width=0.8\textwidth,clip]{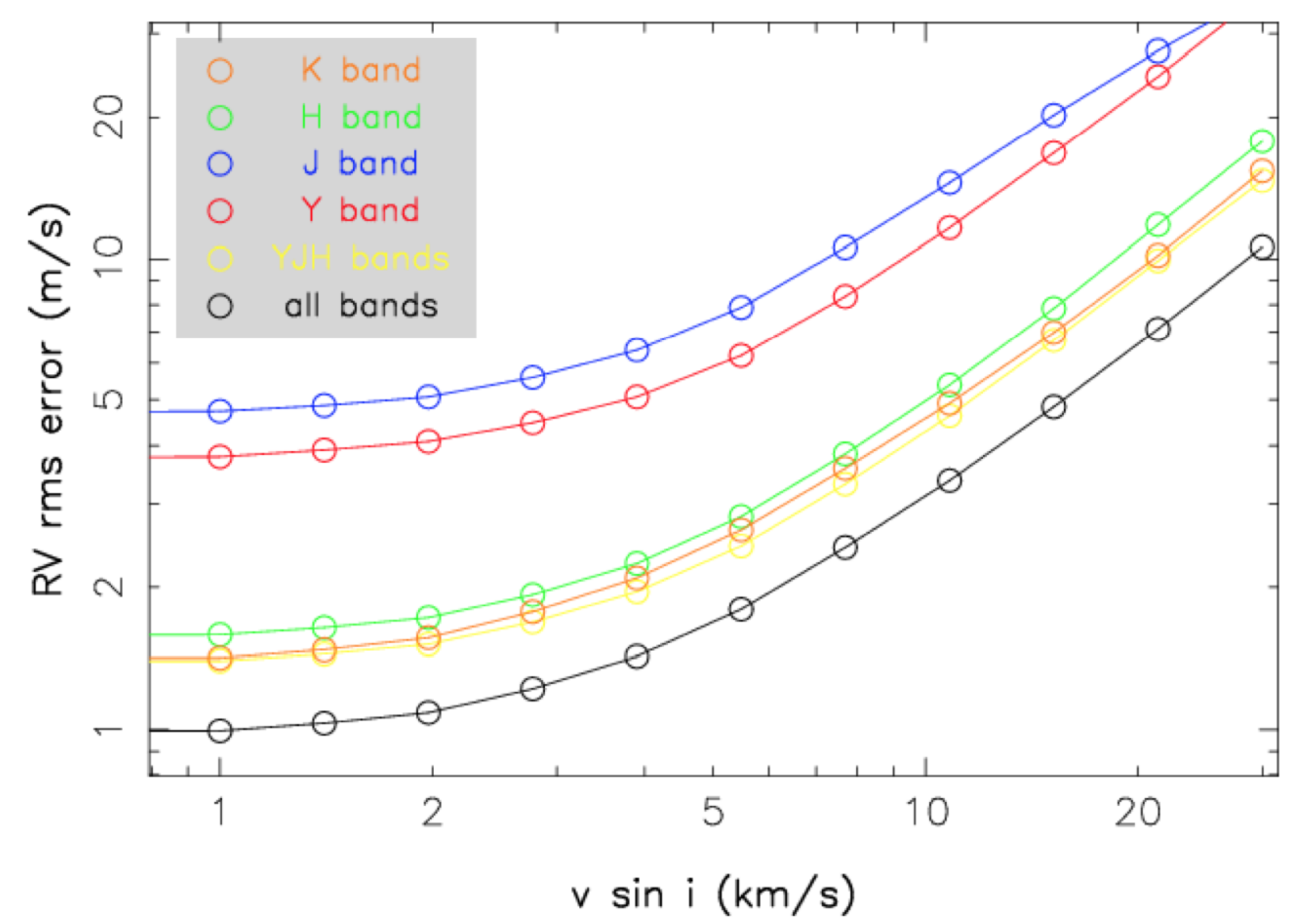}      
  \caption{RV rms photon-noise error (in m/s) vs rotational broadening
    ($v \sin i$) for a M7 dwarf (2,700 K) and for the different nIR bands
    (color curves), assuming a spectral resolution of 75K and a
    peak S/N of 160 per 2 km/s pixel. A RV precision of ~1~m/s can be
    reached at low $v \sin i$'s. The K band (orange) is the main
    contributor to the RV precision and contributes almost as much as
    all other bands (yellow). Without the K band, a twice longer
    exposure time would be required to reach the same RV precision. 
Note that this estimate is likely pessimistic by potentially as much
as a factor of 2, nIR synthetic spectra systematically
under-estimating the strength of many molecular and even atomic
features.} 
  \label{delfosse1:fig1}
\end{figure}

The very-wide simultaneous spectral range, including in particular the
K band, is crucial to SPIRou. It maximizes the instrument efficiency,
both for the exoplanet programs - the K band totaling $\sim$40\% of the RV
content in a full YJHK spectrum for an average M dwarf (see Fig~\ref{delfosse1:fig1}) - and for the
magnetic fields and star/planet formation themes - the relative
spectropolarimetric weight of the K band reaching 60-70\% given the
increase of the Zeeman effect with wavelength as $\lambda^2$. Moreover, the K
band is the only window to access class I embedded protostars, a key
stellar sample to be explored for the first time with SPIRou. The wide
spectral coverage and spectropolarimetric capabilities 
are also unique to SPIRou; other nIR RV instruments currently under
planning (Carmenes on Calar Alto, IRD on Subaru, HZPF on the HET) do
not cover beyond the H band nor include a polarimeter. 
The spectral resolution is also very important, not only to maximize
the velocimetric efficiency, but also to ensure high enough
spectropolarimetric sensitivity to Zeeman signatures; with a spectral
resolution of at least 73.5K, SPIRou is nearing optimal performances.

The resulting instrument concept proposed for SPIRou is a direct
heritage from previous successful instruments built by various members
of the SPIRou project team: HARPS at the 3.6-m ESO telescope
\citep{mayor2003}, ESPaDOnS at CFHT \citep{donati2003} and SOPHIE at 1.93-m OHP
telescope \citep{bouchy2006, bouchy2013}. More specifically, SPIRou includes a
cryogenic high-resolution spectrograph inspired from the evacuated
spectrograph of the HARPS velocimeter, a Cassegrain unit derived from
the ESPaDOnS spectropolarimeter, a fiber-feed evolved from those of
ESPaDOnS and the HARPS/SOPHIE velocimeters, and a Calibration/RV
reference unit largely copied from those of SOPHIE and HARPS.

\subsubsection{The Cassegrain unit}

The Cassegrain unit consists of 2 modules mounted (on top of each
other) at the Cassegrain focus of the telescope. The upper Cassegrain
module includes an ADC correcting the entrance beam for the
atmospheric refraction and a tip-tilt module stabilizing the entrance
image to better than 0.05'' rms; this module also includes a
calibration wheel allowing to inject light from the calibration unit
into the instrument. 
Beginning with a circular instrument aperture of diameter 1.3'', the
lower Cassegrain module mainly includes an achromatic 
polarimeter made of two 3/4-wave dual ZnSe Fresnel rhombs coupled to a
Wollaston prism, splitting the beam into 2 orthogonal
linear-polarization states. The 2 beams emerging
from the beamsplitter are injected into 2 separate fibers at
polarimeter output

\subsubsection{The fiber link and pupil slicer}

The fiber link conveys the light from the twin orthogonally polarized
beams coming out of the Cassegrain polarimeter into the cryogenic
spectrograph. This link consists of a dual 35-m circular fluoride fiber
custom-made with purified material to ensure a throughput of $>$90\%
over the entire spectral range of SPIRou; this fiber link also
includes a pupil-slicer at spectrograph entrance to minimize injection
losses without affecting the spectrograph resolution. The last section of
the fiber link includes a triple 90$\mu$m octogonal fiber (2 for the
science fibers and 1 for a simultaneous RV reference) ensuring a high
scrambling of the near-field image is at least 1000.

\subsubsection{The cryogenic spectrograph}

The high-resolution \'echelle spectrograph is bench-mounted, protected
by one active and 3 passive thermal shields, and enclosed within a
cryogenic dewar. Thanks to a dual-pupil design and an off-the-shelf
commercial R2 \'echelle grating (w/ 23.2~gr/mm), the spectrograph can
record the entire spectral range on a Hawaii 4RG detector (15$\mu$m square
pixels). The pixel size translates into an average spectral bin of
2.28~km/s and the spectrograph features a non-Gaussian instrumental
profile yielding a spectral resolving power of 73.5K. The optical
design of the spectrograph ensure a high total average throughput of
45\% (detector included). The spectrograph is cooled down to 80~K and
thermally stabilized at a rms level of $\sim$2~mK; this thermal stability
ensures in particular that the corresponding spectral drift at
detector level is $<$0.70~m/s on timescales of 1~night. This drift can
be monitored, and thus mostly corrected for, by recording the RV
reference spectrum; in this case, the residual spectral drift at
detector level reduces to $<$0.25~m/s.  

\subsubsection{The calibration and RV reference module}

The calibration module and RV reference unit is used to provide the
instrument with all the required laboratory lamps, and in particular
halogen lamps (for flat fields, used to correct for the detector
pixel-to-pixel sensitivity differences), hollow-cathode lamps
(e.g., Th/U spectra, used to derive the pixel-to-wavelength
calibration relation) and include an additional RV reference, in the
form of a flat field lamp coupled to a Fabry-Perot (FP) etalon,
thermally stabilized at a level of $\sim$10~mK rms to ensure that spectral
lines do not drift by more than 0.25~m/s rms throughout one night. 
This thermalized FP unit (and possibly even the Th/U lamps) could be
replaced in the future with a nIR tunable laser comb (stable to $<$0.10~m/s) 
to further improve the overall RV precision of the instrument. 

\subsection{The international project team}

The SPIRou project team gathers a number of partners from different
institutes and countries. 
More specifically, the team includes several institutes from France
(IRAP and OMP in Toulouse, IPAG in Grenoble, OHP and LAM in Marseille,
plus an extended science team from IAP / LESIA / CEA / LERMA / IAS /
LUTH / LATMOS based in Paris and surroundings), from Canada (UdeM / UL
in Montr\'eal and Qu\'ebec City, NRC in Victoria), from Switzerland
(Geneva Observatory), from Taiwan (ASIAA in Taipei), from Brazil
(LNA in Itajuba, plus additional science contribution from UFRN / UFMG
in Natal and Belo Horizonte), from Portugal (CAUP in Porto) and from
CFHT.

\section{Main science goal \#1 : exoplanets around very-low-mass stars}
 
\subsection{Scientific context}

One of the 2 main goals of SPIRou is to search for, and to
characterize, exo-Earths orbiting low-mass stars - with a particular
interest for planets located in the habitable zone (HZ) of their host
stars. The study of exoplanetary systems is one of the most exciting
areas of astronomy today. Identifying habitable Earth-like planets and
searching for biomarkers in their atmospheres is among the main
objectives of this new century's astronomy, motivating ambitious space
missions (e.g., JWST, TESS, CHEOPS, EChO, PLATO). Among the various
techniques developed to detect exoplanets, two are very efficient and
complementary.  Whereas RV studies look for Doppler
shifts induced by orbiting planets in the spectrum of their host
stars, giving access to the planet mass, long-term photometric
monitoring searches for regular occultations caused by planets
transiting the visible stellar disc, yielding the planet radius. For
exoplanets detected with both techniques, one can estimate their
densities and thus constrain their bulk compositions. Provided host
stars are bright enough, one can even probe the outer atmosphere of
transiting planets using transit spectroscopy, opening the new
research field of exoplanetology \citep{charbonneau2007}. 

In this context, much interest has recently been focused on low-mass M
dwarfs, around which habitable super-Earths are much easier to
detect. To be considered potentially habitable, planets must be within
the proper range of orbital distances where liquid water can be stable
on their surface. This constraint also imposes limits on the
atmospheric pressure at the planet surface, and thus indirectly on the
planet mass. The range of orbital distances for HZs also strongly
depends on the mass (and thus on the temperature) of the host star,
with lower temperatures moving HZs closer in. Habitable
exo-Earths around M dwarfs are thus expected to produce much larger
RV wobbles (4$\times$ to 8$\times$ for M4 and M6 dwarfs, respectively) compared to
the same planet orbiting a Sun-like star. A 1~m/s RV precision is
sufficient to detect habitable telluric planets around M dwarfs - the
much shorter orbital periods (of order of weeks) vastly decreasing the
timescale over which observations must be collected; this is how the
first likely- habitables super-Earths were discovered \citep{udry2007,
  mayor2009, delfosse2013, bonfils2013a}.  

Photometric transits are also much deeper for M dwarfs as a result of
their smaller radii - by 11$\times$ and 45$\times$ for M4 and M6 dwarfs,
respectively. A prime goal of the coming years is to discover Earths
or super-Earths whose atmosphere can be scrutinized and characterized
with space missions (such as JWST and/or EChO) in the next
decade. Since atmospheric characterization primarily requires as deep
an atmospheric transit as possible on the one hand, and as bright a
star as possible on the other hand (in the nIR, where absorption from
atmospheric molecules mostly concentrates), M dwarfs are optimal
targets for this quest \citep{rauer2011}. Today, only a
handful of very-bright transiting systems have been discovered up to
now - most being giant gaseous planets - but many more are expected
with forthcoming space missions like TESS or PLATO. 

Last but not least, statistical properties of planets around M dwarfs
(compared to those around Sun-like stars) can provide key information
on planetary formation, and in particular on the sensitivity of planet
formation to initial conditions in the protoplanetary disc
\citep[e.g.,][]{ida2005}. That M dwarfs vastly dominate the stellar 
population in the solar neighborhood and are likely hosting most
planets in our Galaxy only makes this study even more crucial. 

\subsection{The SPIRou planet search}

\begin{figure}[ht!]
 \centering
 \includegraphics[width=0.8\textwidth,clip]{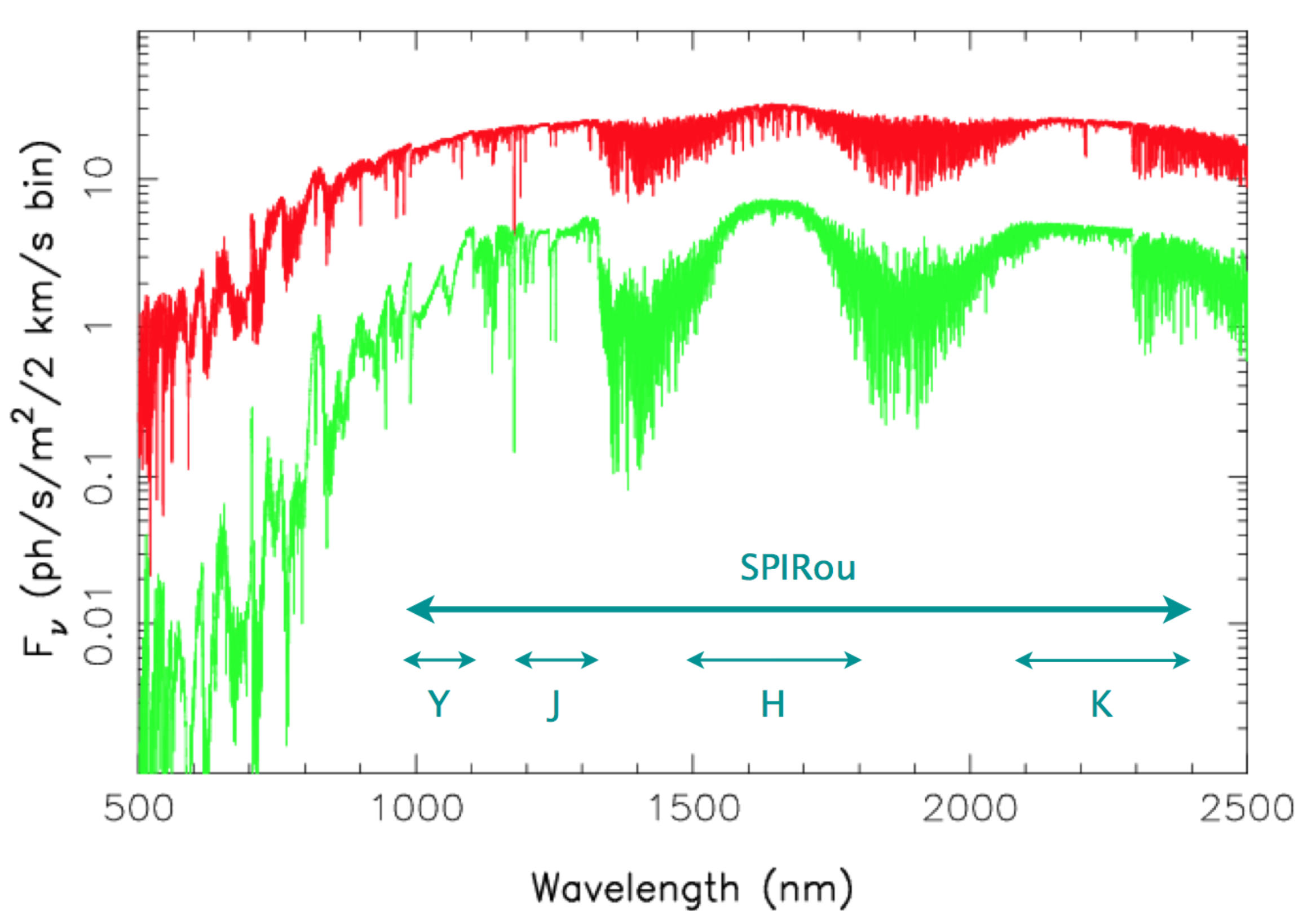}      
  \caption{Photon distribution (per 2~km/s velocity bin size) for a M6
    (3,100~K, red) and M8 (2,300~K, green) dwarfs at 10~pc (derived
    from NextGen models \citep{allard1997}. M6 and M8 dwarfs
    respectively produce $\sim$30 and $\sim$1000 times more photons (per
    velocity bin) in HK than in V.}  
  \label{delfosse1:fig2}
\end{figure}

Based on high-precision RV measurements, the SPIRou planet search we
propose will greatly expand the current exploratory studies carried
out with existing visible velocimeters (e.g., HARPS@ESO, SOPHIE@OHP)
by giving access to a large sample of stars inaccessible with existing
instrumentation. The SPIRou planet search will in particular build
upon the success of the pioneering HARPS RV survey of M dwarfs 
\citep{bonfils2013b}, which demonstrated that super-Earths with orbital periods $<$100~d 
are more numerous around M dwarfs than around
Sun-like stars, with an occurrence frequency close to 90\%; moreover,
preliminary results suggest that
about half of these super-Earths are located in the HZs of their host
stars. With existing velocimeters such as HARPS, RV measurements with
a precision of 1~m/s are possible for only the $\sim$100 brightest M
dwarfs. This is clearly insufficient, either to have a realistic
chance of detecting several transiting habitable super-Earths or to
achieve a proper statistical survey of rocky exoplanets around M
dwarfs. Given their low temperatures, red and brown dwarfs are much
more accessible at nIR wavelengths (see Fig~\ref{delfosse1:fig2}). In addition to a 1~m/s RV
precision and a high throughput, SPIRou offers the widest simultaneous
nIR spectral coverage (0.98-2.35~$\mu$m) yet available on any
telescope, making it optimally suited for carrying out efficient, 
systematic RV exoplanet surveys of M dwarfs.

SPIRou will also crucially contribute to the forthcoming extensive
photometric surveys of transiting planets around M dwarfs, either from
space (e.g., TESS, CHEOPS, PLATO) or from the ground (e.g.,
ExTrA). Spectroscopy is indeed mandatory to discard false detections
(e.g., background eclipsing binaries), to establish the planetary
nature of all transiting objects detected around low-mass dwarfs
through photometric monitoring and to measure their mass from RV
measurements. A high-precision velocimeter working in the nIR will
thus be essential to monitor all candidates detected with ground and
space photometers around bright M dwarfs, and in particular around
late-M ones, hardly accessible to velocimeters working in the
visible. A nIR spectrograph will also usefully contribute to the quest
for close-in transiting exo-Earths around bright M dwarfs through a
systematic survey prior to any photometric observations.

More specifically, SPIRou will contribute to exoplanet science along 3
main avenues, that we foresee as the prime exoplanet themes of the
SPIRou planet search.

\subsubsection{Follow-up of transiting planet candidates uncovered by future photometric surveys}

Among the 3,500+ planet candidates yet found by
Kepler, only a few tens have been validated and characterized through RV measurements,
the vast majority of candidates orbiting stars that are too faint for
current RV surveys \citep[see][for more details]{santerne2013}. The
goal of future photometric surveys is to
detect planet candidates around brighter stars, with a specific
emphasis on nearby M dwarfs. Among them, the TESS space mission, to be
launched in 2017 and predicted to detect $\sim$300 super-Earths, is
certainly very promising. Since (i) most Earths and super-Earths
detected with TESS will orbit around M dwarfs, and (ii) less than
$\sim$30\% of them will be accessible to optical RV follow-ups
\citep{deming2009}, SPIRou will be the best RV instrument to
monitor in the nIR the $\sim$150 best candidates visible from CFHT, to
confirm or reject their planetary nature and to determine their
masses.

Monte Carlo simulation show that with $\sim$60 visits per star and
with S/N$\sim$160 spectra per visit SPIRou has the capacity to
validate and characterize planets of Earth Mass, orbiting mid-M dwarfs with a period of
$\sim$30~d. This observational effort requires a total of 150 CFHT  
nights. 

\subsubsection{RV survey of a large sample of M-dwarfs}

\begin{figure}[ht!]
 \centering
 \includegraphics[width=0.8\textwidth,clip]{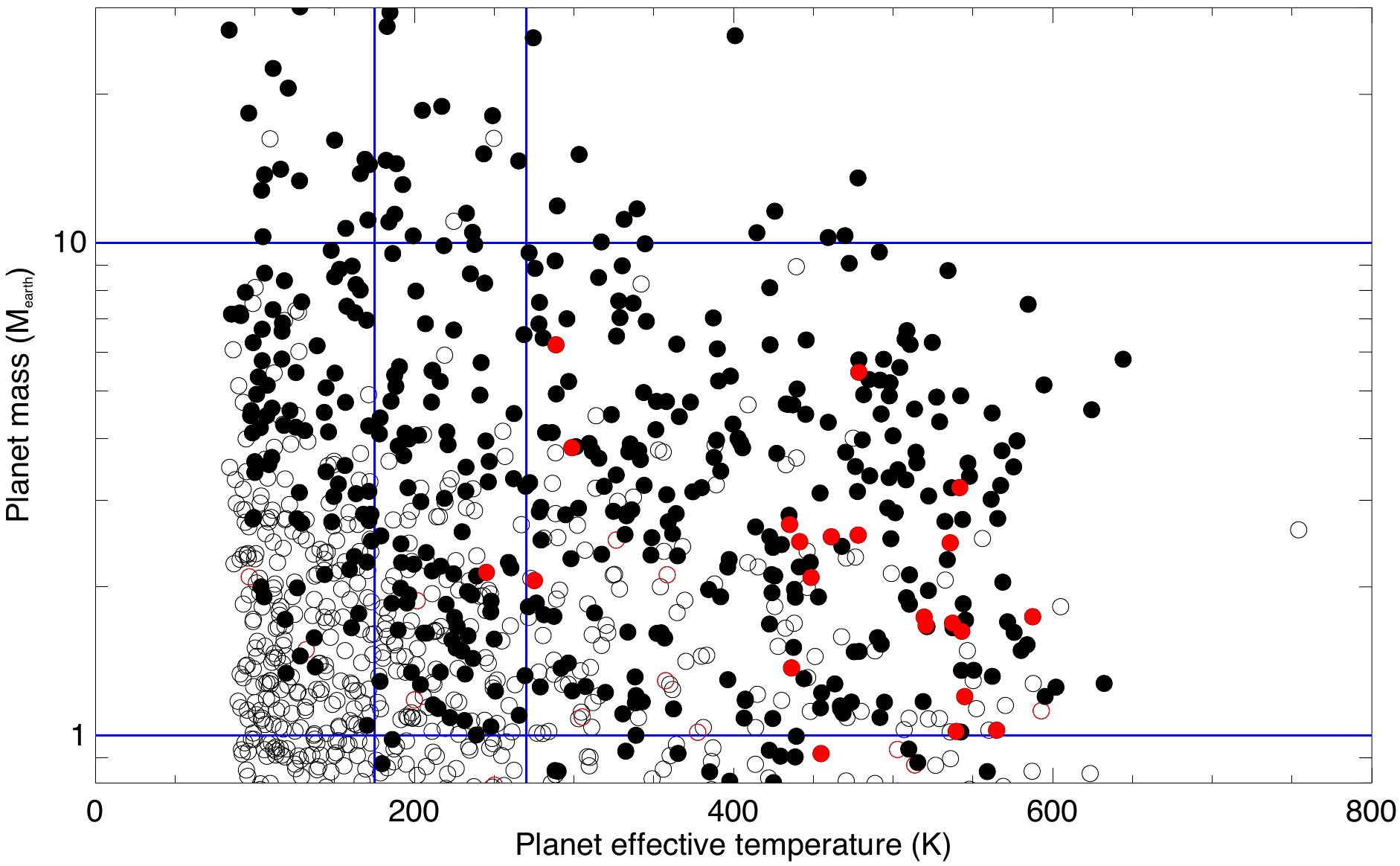}      
  \caption{Planets found with a 600-target survey according to our
    Monte Carlo simulation. Filled circled
    indicate detected planets, open circles undetected ones and red
    circles (both filled and open) represent transiting planets. Blue
    lines show notional limits for the habitable zone, both in mass
    and temperature. Most planets with $>$2~M$_{\oplus}$ in the habitable zone are
    detected, including one transiting. Interestingly, a sample of
    sub-M$_{\oplus}$ planets with T$_{\rm eq}>$350 K is also detected. 
}  
  \label{delfosse1:fig3}
\end{figure}

As TESS will majoritarly operate on 27-d
windows of continuous monitoring for most stars, the majority of
planet candidates showing at least 2 transits will have periods $<$20~d
and will not be located in the HZ of their host stars. For planets
with longer periods, and in particular for those located in the HZ,
RV-driven planet searches will be more efficient. Our Monte Carlo simulations
demonstrate that (see Fig~\ref{delfosse1:fig3}), with a survey
focussing on $\sim$600 M dwarfs (requiring 
600 CFHT nights for $>$60 visits per stars), SPIRou could potentially
detect $\sim$450 new exoplanets, 
$\sim$300 being less massive than 5~M$_{\oplus}$; among the latter sample, $\sim$50 would
be orbiting in the HZ and $\sim$15 would be transiting, while $\sim$2 would have
both characteristics. This survey should
allow to determine ${\eta}_{\oplus}$, the faction of habitable planets in the Solar
neighborhood, with an accuracy of $<$10\%. Photometric follow-ups of all
planets detected with SPIRou will be achieved, e.g., with CHEOPS and
ExTrA, to determine which ones are transiting, once their ephemeris
and transit windows are well known. Identifying transiting habitable
super Earths is crucial for all future attempts at detecting
biomarkers in their atmospheres with JWST or EChO. 

\subsubsection{Occurrence frequency of planets around M dwarfs}

By expanding the
sample by 10$\times$ (with respect to the existing optical surveys of M
dwarfs) and thus by bringing a 3 fold improvement in the statistics
of planet properties, the SPIRou observations outlined in the 2 first
items of our planet search will provide much more reliable constraints
on planet formation models. Moreover, by extending the RV monitoring
on a selected sample of M dwarfs and on a larger time span, SPIRou will
likely reveal additional bodies in most systems at larger period. This extended
monitoring, not included in the first part of RV of our planet search,
will be carried out on the $\sim$350 most interesting M dwarfs with
detected planets / systems and will require an additional amount of
250 CFHT nights to achieve 40 more visits per star.

\section{Main science goal \#2 : magnetic fields and star / planet Formation}

\subsection{Scientific context}

The other main goal of SPIRou is to explore the impact of magnetic
fields on star and planet formation, by detecting and characterizing
magnetic fields of various types of young stellar objects (e.g.,
classical T Tauri stars, embedded class-I protostars, young
protostellar accretion discs). This quest
will expand the pioneering surveys carried out in the framework of the
study with optical spectropolarimeter, mainly ESPaDOnS@CFHT. 

Studying how Sun-like stars and their planetary systems form comes as
a logical addition to the direct observation of exoplanets. Within the
last decades, this research field underwent major observational and
theoretical advances, for instance by clarifying the crucial role of
magnetic fields, not only on the gravitational collapse of giant
molecular clouds \citep[e.g.,][]{hennebelle2008a}, but
also on the formation of accretion discs and pre-stellar cores
\citep[e.g.,][]{hennebelle2008b} from which stars and their
planetary systems are born.     

At an age of $\sim$1~Myr, low-mass protostars emerge from their dust
cocoons, most often surrounded by a massive accretion disc in which
planet form and migrate. This is the so-called ``classical T Tauri''
(cTTS or class-II protostar) stage - one of the best studied phase of
stellar formation thanks to its relative accessibility to existing
instruments. Observations suggest in particular that magnetic fields
of cTTSs are strong enough (i) to disrupt the central regions of the
surrounding accretion discs, thereby generating magnetospheric gaps at
the heart of the discs, (ii) to guide the plasma from the discs to the
stars along discrete magnetospheric accretion funnels, and (iii) to
drastically slow-down their rotation rates by magnetically coupling
stellar surfaces with the inner edges of the accretion discs
\citep[e.g.,][]{bouvier2007}. 

Spectropolarimetric observations secured with ESPaDOnS@CFHT enabled to
disclose, for a small sample 
of $\sim$15~cTTSs, the large-scale magnetic topologies that link low-mass
protostars to their accretion discs, and to demonstrate that this
topology strongly relates to the internal structure of the protostar,
and thus to both its age and mass \citep[e.g.,][]{donati2010}. When the
protostar is young enough and has a low-enough mass to 
be fully-convective, its large-scale magnetic topology is dominated
by a strong dipolar-like field roughly aligned with the stellar
rotation axis - thereby providing a quantitative explanation of the
physical star/disc coupling mechanism through which the protostar is
strongly spun down \citep[e.g.,][]{zanni2013}. These
observations are however still rather sparse as a result of the
relative faintness of cTTSs (at optical wavelengths); moreover,
younger class-I protostars (with ages $<$1~Myr), for which magnetic
fields are expected to have an even bigger evolutionary impact, are
still out of reach of existing instruments, their dust cocoon hiding
them completely from view at optical wavelengths.

\subsection{The SPIRou survey of magnetic protostars}

\begin{figure}[ht!]
 \centering
 \includegraphics[width=0.8\textwidth,clip]{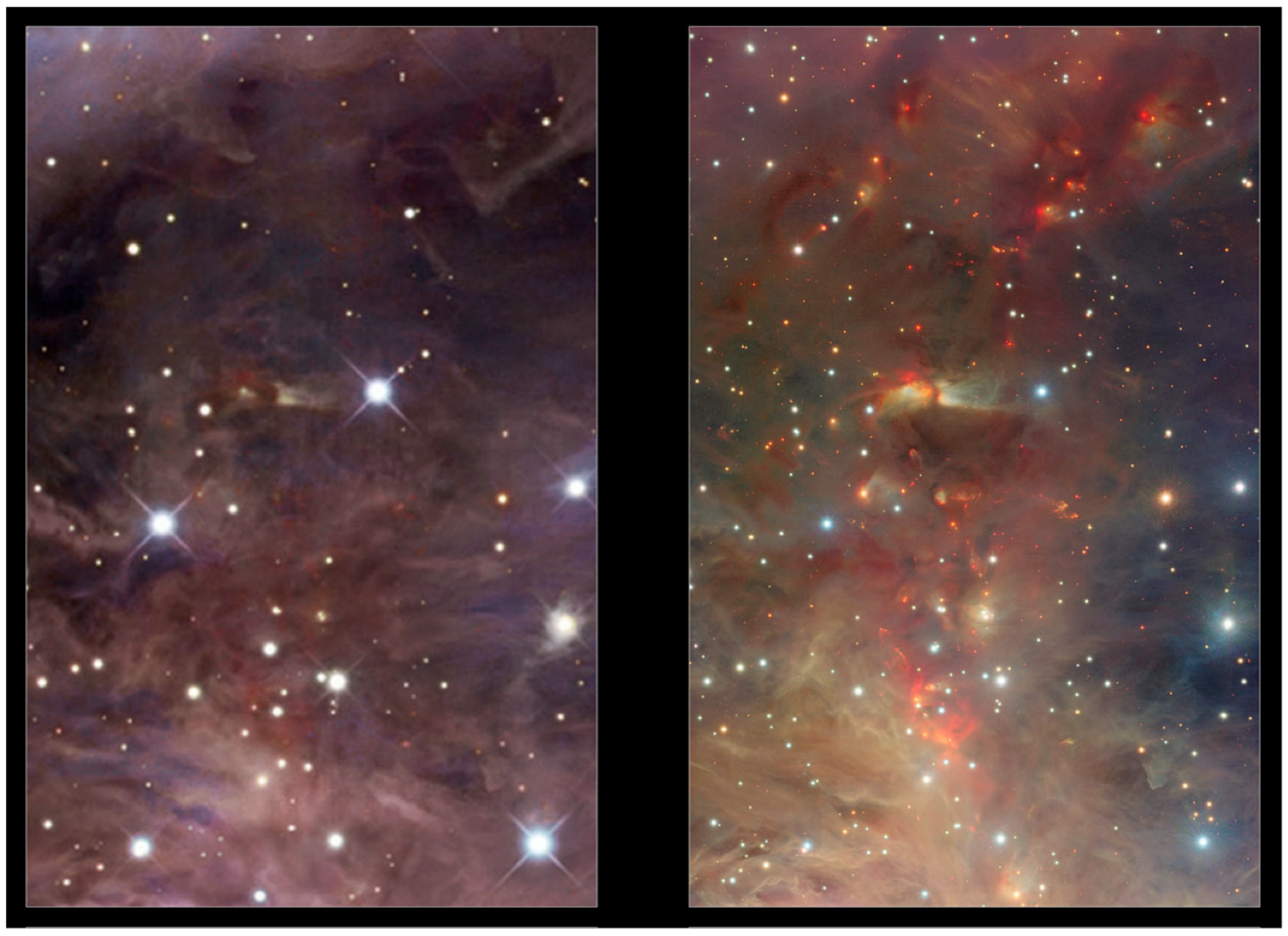}      
  \caption{The Orion nebula as seen in visible (left) and IR (right)
    light (ESO/VISTA). In the IR, dust cocoons around young stars
    are much more transparent. }  
  \label{delfosse1:fig4}
\end{figure}

With a much higher magnetic sensitivity than ESPaDOnS, thanks to
both the increased nIR brightness of protostars (especially in K, see
Fig~\ref{delfosse1:fig4}) and the enhanced Zeeman effect at larger wavelengths, SPIRou
will provide a much deeper and more systematic access to large-scale
fields of class-I and -II protostars. More specifically, it will allow
(i) to survey a 5-10$\times$ larger sample of cTTSs than the very limited one
currently accessible with ESPaDOnS, and (ii) to extend for the first
time this study to the brightest class-I protostars thanks to the K band coverage.
The suggested SPIRou survey will ideally complement ALMA observations of pre-stellar (class-0) cores 
and of their magnetic fields, and will thus bring one of the key missing pieces in our
understanding of star / planet formation.

SPIRou will also have the power to detect hot Jupiters orbiting
around more-evolved class-III protostars (the so-called ``weak-line T
Tauri'' stars or wTTSs) and thus to verify whether close-in giant
planets are either much more or much less frequent around low-mass
protostars than around mature, Gyr-old Sun-like stars. These
observations will thus yield a direct observational test of the
formation and migration of hot Jupiters, allowing to estimate the
relative fraction produced through disc migration (acting during the
formation stage) and that attributable to interactions / scattering
(occurring much later). Finally, SPIRou will also be able to observe
the innermost regions of protostellar accretion discs, out of reach of
ALMA, to detect and characterize their magnetic fields and to identify
the potential presence of migrating hot Jupiters
\citep[e.g.,][]{donati2005,powell2012}.

\begin{figure}[ht!]
 \centering
 \includegraphics[width=0.8\textwidth,clip]{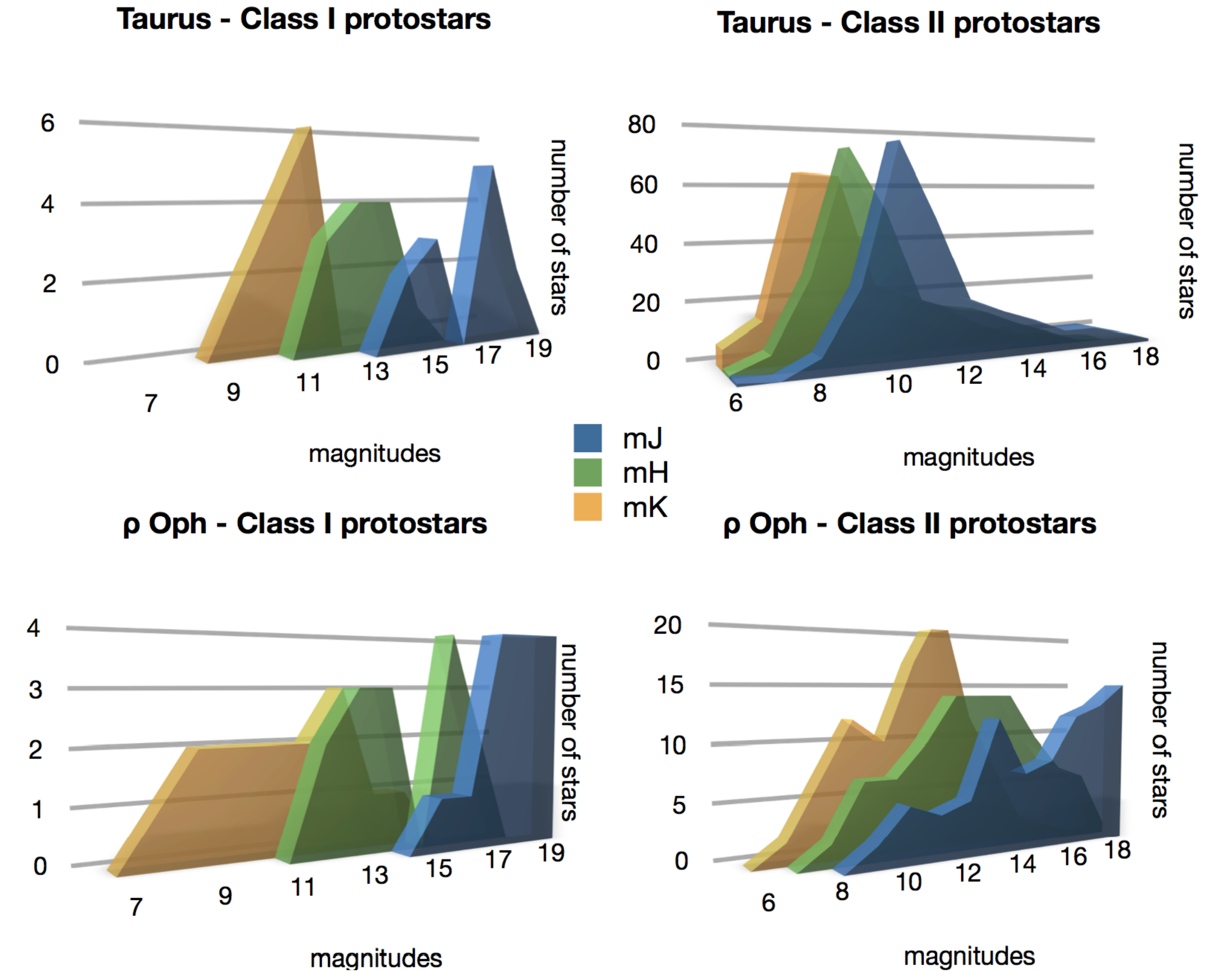}      
  \caption{JHK magnitude histograms of class-I (left) \& -II (right)
    protostars in Taurus (up) and $\rho$~Oph (bottom). Class-I
    protostars are mostly brighter than 12 in K only (as a result of
    obscuration, especially in $\rho$~Oph), whereas a significant
    fraction of class-II protostars are accessible in all 3 bands. 
 }  
  \label{delfosse1:fig5}
\end{figure}

With a survey carried out on 5 of the most accessible star forming
regions (e.g., Taurus/Auriga, TW Hya Association, $\rho$ Ophiuchius,
Lupus, Orion Nebula Cluster; see Fig~\ref{delfosse1:fig5}), SPIRou can detect for the first time the
large-scale fields of $\sim$50 embedded class-I protostars, bringing yet
unknown information on how magnetospheric accretion operates at so
early a step in the formation process. In addition to this, SPIRou
will be able to monitor $\sim$200 class-II and -III protostars (cTTSs and
wTTSs), expanding the pioneering ESPaDOnS survey by 5-10$\times$ into
full-scale surveys of magnetic protostars and their close-in giant
planets. This survey will require a total of 250 CFHT nights; on top
of this, monitoring a small sample of $\sim$10 protostellar accretion discs
will require an additional 50 nights. 

\section{Additional science goals}

In addition to the two main goals, SPIRou will be a
very innovative and efficient instrument for tackling many more
science themes. A few of them are briefly outlined below.

\subsection{Large-scale dynamos of M-dwarfs}

Using the spectropolarimetric data collected for the exoplanet survey
of M dwarfs, SPIRou can also study the large-scale dynamo
fields of fully convective dwarfs. These magnetic fields are indeed
the main source of their activity and therefore a potential drawback
for the habitability of their planets \citep{lecavellier2012,vidotto2013};
studying dynamos of fully convective bodies can also be very
informative on magnetic fields of Earth-like exoplanets 
\citep[e.g.,][]{christensen2009,reiners2010} and usefully complement direct sensitive radio
observations \citep[e.g., LOFAR,][]{zarka2010}, with the ultimate
aim of working out whether magnetic fields of exoplanets can improve
their habitability.

Published studies of large-scale fields of fully-convective dwarfs
have already demonstrated that these magnetic topologies are very
sensitive to the aspect ratio of the convective zone and even suggest,
for very-low mass dwarfs, a bistable behavior of the underlying dynamo
processes potentially similar to that invoked for planetary dynamos
\citep{morin2008,morin2010,morin2011}. By comparing magnetic topologies of
M dwarfs with theoretical predictions and results of numerical
simulations, and by trying to generalize these results to planetary
dynamos, one should ultimately be able to better understand the
physical processes capable of amplifying and sustaining large-scale
magnetic fields in both fully-convective dwarfs and planets 
\citep[e.g., ][]{schrinner2012}. Using data from the exoplanet
survey, SPIRou will provide a thorough census of magnetic topologies
of M dwarfs (and in particular of fully-convective ones) that will
usefully guide theorists towards more realistic, generalized dynamo
models in better agreement with observations of both stellar and
planetary large-scale fields.

\subsection{Studies of planetary atmospheres}

SPIRou can also very efficiently contribute to atmospheric studies
of telluric or giant planets, whether or not they belong to the solar system. 

In the case of the solar system, SPIRou will be able to carry out
spatially-resolved, detailed spectroscopic studies of the chemical
composition (at both low and high altitudes), of the wind dynamics and
of the auroral emission of planetary atmospheres. These studies will
allow in particular to better understand the complex interactions
between atmospheric volatiles, planetary interiors, surfaces and
climates \citep[e.g.,][]{bezard2009}; they can also
accurately estimate wind velocities at different atmospheric locations
and altitudes, as well as their temporal variability \citep[e.g.,][]{widemann}. 
Finally, auroral emission (and polarization) can inform on potential links between planetary
atmospheres and magnetospheres. Thanks to its wide spectral domain
(including the K band), to its high RV precision and to its
spectropolarimetric capabilities, SPIRou will be able to very
significantly contribute to chemical and dynamical studies of
solar-system planet atmospheres. 

Exoplanet atmospheres are obviously much more elusive and tricky to
detect and to characterize. In the particular case of transiting
exoplanets, atmospheres can be scrutinized either by transmission
during a planetary transit, or by occultation during a planetary
eclipse. For close-in planets (and in particular hot Jupiters), it
could be possible to detect from the ground the spectral contribution of
the star-lit side of the planet, by monitoring the Doppler shift
(induced by the planet orbital motion) of specific atmospheric species
\citep[e.g., CO,][]{snellen2010}. SPIRou will thus be
able to contribute to this quest in a original way,
thanks to its wide spectral domain
and to the K band in particular (that includes a number of key
atmospheric markers).
 
\subsection{Weather patterns of brown dwarfs} 

Studying the atmospheres of ultra-cool L and T brown dwarfs (BDs) is
yet another obvious research field for SPIRou. Despite huge modeling
efforts invested since the discovery of BDs, many questions remain
open - likely related to the complex physics of BDs atmospheres and in
particular to mechanisms of dust clearing through specific weather
patterns occurring in their atmospheres \citep[e.g.,][]{radigan2012}. 
The disc-averaged spectral energy distribution (SED) of
ultra-cool BDs may indeed not be representative of any single region,
and thus cannot be modeled using a unique set of physical parameters
(e.g., temperature, dust-settling, grain properties). Evidence that
this is likely the case comes from the fact that ultra-cool BDs often
exhibit photometric variability at a level of a few \% up to a
remarkable 25\% \citep{artigau2009}. This variability
is apparently due to a combination of rotational modulation and
intrinsic evolution on short timescales, likely caused by weather-like
clearings in the dust- cloud deck \citep[e.g.,][]{littlefair2006}. 
Unravelling the physics of these weather patterns and
distinguishing between the several theoretical scenarios requires
observations capable of localizing the dust clouds in BD atmospheres
and following their rapid evolution with time. 

Doppler imaging through high-resolution nIR spectroscopy is a very
attractive and viable approach to map weather patterns of BDs; most
BDs are indeed rapid rotators and often exhibit rotational modulation,
both photometrically and spectroscopically, providing ideal conditions
for Doppler imaging. Though not yet applied to stars cooler than mid-M
due to their intrinsic faintness at optical wavelengths, Doppler
imaging of BDs in the nIR is perfectly feasible with SPIRou, opening a
new window for studying chemical inhomogeneities in their
atmospheres. SPIRou will be able to monitor $\sim$10 L and T BDs among the
best suited for this experiment, for a total amount of 30 nights.

\section{The SPIRou Legacy Survey }
 
The amount of observing time required to complete the
two main science goal is large (1300 nights). In this context,
SPIRou only makes sense if coupled to a SPIRou Legacy Survey
of 500 CFHT nights on a timescale of $\sim$5~yr focussed mostly 
on both main science goals.
The current plan is to divide the SPIRou Legacy Survey into three main components, 
two of them being dedicated to exoplanets around M dwarfs, while one will be concentrating 
on large-scale magnetic fields and young hot Jupiters of class-I-III protostars.

\section{Calendar}

SPIRou has successfully passed the preliminary design phase (PDR) in
October 2012 and is now in the final design phase.  Provided SPIRou is selected by CFHT and
succeeds the upcoming final design review (FDR), it should be
installed on the 3.6-m CFH telescope in early 2017 after the following phases :

\begin{itemize}
\item early 2014: design validation (FDR) 
\item 2014-2015: construction and acceptance tests of all individual SPIRou subsystems
\item 2016: instrument integration and acceptance tests at IRAP 
\item 2017: installation at CFHT, first light, technical commissioning and science verification
\end{itemize}

We further stress that, with this schedule, SPIRou is ideally phased with the predicted launch times of 
both TESS and CHEOPS (2017) as well as that of JWST (2018).

\bibliographystyle{aa}  
\bibliography{spirou_main} 

\end{document}